\documentclass[11pt, a4paper,fleqn]{article}
\usepackage{amssymb}
\usepackage{amsmath}
\usepackage{graphics}
\usepackage{graphicx}
\usepackage{bm}
\usepackage{xcolor}
\usepackage{fdsymbol} 
\usepackage{makecell}
\usepackage{pdflscape}
\usepackage[round,authoryear]{natbib}
\usepackage{authblk} 
\usepackage{rotating}
\usepackage{multirow}
\usepackage{ulem}
\usepackage{listings} 
\usepackage{inconsolata} 
\numberwithin{equation}{section}

\begin{document}
\title{Exact matching as an alternative to propensity score matching}
\author[1,2]{Ekkehard Glimm\footnote{email: ekkehard.glimm@novartis.com}}
\author[1]{Lillian Yau}
 
\affil[1]{Novartis Pharma AG, Basel, Switzerland}
\affil[2]{University of Magdeburg, Magdeburg, Germany}

\maketitle
\vspace*{-0.5cm}

\section*{Abstract}\label{abstra}

The comparison of different medical treatments from observational studies or across different clinical studies is often biased by confounding factors such as systematic differences in patient demographics or in the inclusion criteria for the trials. Propensity score matching is a popular method to adjust for such confounding. It compares weighted averages of patient responses. The weights are calculated from logistic regression models with the intention to reduce differences between the confounders in the treatment groups. However, the groups are only ``roughly matched" with no generally accepted principle to determine when a match is ``good enough”. 

In this manuscript, we propose an alternative approach to the matching problem by considering it as a constrained optimization problem. We investigate the conditions for exact matching in the sense that the average values of confounders are identical in the treatment groups after matching.  
Our approach is similar to the matching-adjusted indirect comparison approach by Signorovitch et al. (2010) but with two major differences: First, we do not impose any specific functional form on the matching weights; second, the proposed approach can be applied to individual patient data from several treatment groups as well as to a mix of individual patient and aggregated data.

\section{Introduction}\label{intro}

Clinical trials are designed to investigate the effectiveness of different treatment options for a disease. In an ideal world, we would like to calculate individual treatment effects for every patient, but since it is impossible to treat anyone with two or more different treatment options at the same time, 
this ideal is unachievable. Thus, rather than comparing treatment effects on every individual directly, we compare groups of patients who receive different treatments. If treatment effects are different for different patients, we must resort to comparisons of population-average effects. In order to isolate treatment effects from confounding inferences, it is then of crucial importance that the groups of differently treated patients are not different with respect to any confounder.


In randomized controlled clinical trials, randomization guarantees that any potential confounding affects the compared groups in the same way. For example, if the disease stage has an influence on treatment efficacy, randomization makes sure that the distribution of disease stage categories is the same in the compared treatment arms. When analyzing observational data, however, non-randomized treatment groups are compared. Thus, a substitute for randomization is needed.

Many statistical techniques have been developed to adjust for confounding when comparing data collected from observation studies, see \citet{hernan2023} for a comprehensive overview. Propensity score matching is one of the most popular approaches. When comparing two studies, it weights the patients in two populations (e.g. patients from two different clinical studies) in such a way that they are rendered matched and hence comparable.
The notion of ``comparable", however, remains somewhat vague when applying this approach. After weighting, the weighted averages of patient characteristics which are deemed to be potential confounders of treatment effect are closer than before, but they are not necessarily identical. This has lead to some discussion regarding quantification of ``closeness" \citep{franklin2014}. In this paper, we explore a direct standardization that gives exact matching as an alternative avoiding the issue entirely.

The paper is organized as follows: In section \ref{motivation}, we introduce three simulated examples to illustrate concepts and emphasize certain properties of the suggested approaches. Two real datasets from clinical studies are also introduced. Section \ref{matching} discusses propensity score matching and contrasts it with the suggested direct standardization which results in exact matching. Section \ref{results} applies both approaches to the examples from section \ref{motivation}. Section \ref{simstudies} presents a simulation study of 1,000 pairs of simulated IPD's. Section \ref{statcon} provides statistical interpretations of the exact matching technique by direct standardization, especially for the outcome responses. We conclude with a discussion in section \ref{discussion}.

\section{Motivation}\label{motivation}

\subsection{Illustrative simulated data}\label{sub_illuexp}

We simulate three examples. Each example consists of two studies. They both contain individual patient level data (IPD). We label the two studies in each example as IPD A and IPD B. Each IPD has two continuous variables, $x_1$ and $x_2$, which are used in matching. Two-dimensional data is used here for easy visualisation. In reality, matching is usually conducted with high-dimensional data that can have both continuous and categorical variables.

\begin{figure}
    \centerline{\includegraphics[width=1.2\textwidth]{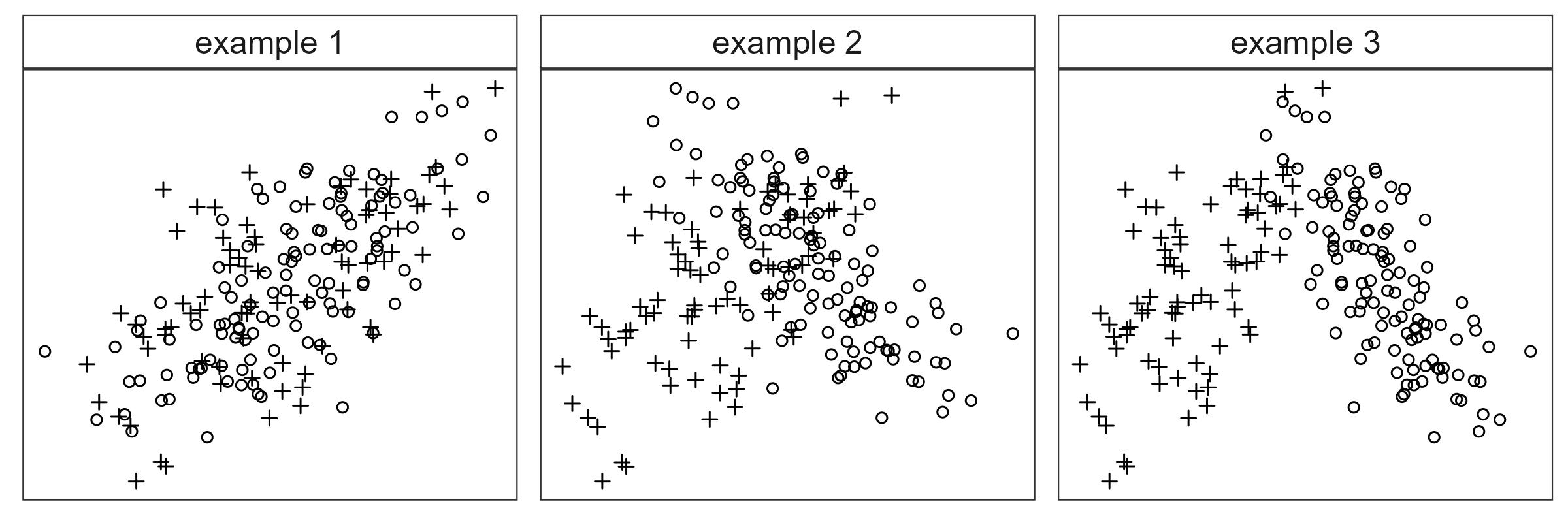}}
    \centering\footnotesize{+ IPD A $\quad \quad \circ$ IPD B\\ $x_1$: horizontal axis; $x_2$: vertical axis}
    \caption{Three pairs of simulated IPD's}
    \label{fig:plt0}
\end{figure}

Figure \ref{fig:plt0} presents three scatter plots for the three examples. In each plot, observations from IPD A are indicated with $+$'s and those from IPD B with $\circ$'s. Variable $x_1$ is plotted along the horizontal axis and $x_2$ along the vertical axis. The means of the variables are given in Table \ref{table1}. 

\begin{table}[h!]
    \centering
    \begin{tabular}{c|cc|cc|cc} \hline
    & \multicolumn{2}{c|}{example 1} & \multicolumn{2}{c|}{example 2} & \multicolumn{2}{c}{example 3} \\
    & $\bar{x}_1$ & $\bar{x}_2$ & $\bar{x}_1$ & $\bar{x}_2$ & $\bar{x}_1$ & $\bar{x}_2$ \\ \hline
       IPD A & -0.18 & -0.12 & -0.18 & -0.12 & -0.18 & -0.12 \\
       IPD B & -0.05 & -0.05 & 1.05 & 0.45 & 3.08 & -0.05 \\ \hline 
       Combined & -0.12 & -0.08 & 0.43 & 0.17 & 1.45 & -0.08 \\ \hline
    \end{tabular}
    \caption{Summary statistics}
    \label{table1}
\end{table}

In the first example, both IPD's overlap considerably, indicating that the two sets of study participants are sampled from a common underlying patient population. In the second example, the correlation between the two variables are in completely different directions. Nonetheless, there is an intersection of patients who fit the profiles of both populations. In the third example, the two IPD's have least amount of overlap, which suggests the participants are from very different populations.

We use these three artificial examples to illustrate the method. Results are presented in Section \ref{results_illu}.

\subsection{Real data}\label{sub_real}

\citet{hampson2024} demonstrated the use of a target trial framework when comparing a single arm clinical trial with an external control arm. The application compared ELARA, a single arm clinical study of a CAR-T therapy for patients with relapse or refractory follicular lymphoma, to an external control based on data collected from a non-interventional retrospective chart review study, titled ReCORD, of patients with the same disease condition but only receiving standard of care. To emulate randomization and adjust for potential confounding, baseline covariates for patient $i$ in ReCORD were weighted with the odds $p_i/(1-p_i)$, where $p_i$ is the propensity score, i.e. the estimated conditional probability that patient $i$ would have been assigned to receive CAR-T therapy given their baseline covariates. Efficacy outcomes were then compared between ELARA and the weighted ReCORD cohorts.

In this article, we utilize the baseline covariates from the two studies, and demonstrate the use of exact matching by direct standardization as an alternative to the propensity odds weights. 

Observed means along with results of the standardized means are presented in Section \ref{results_real}. For comparison the propensity score weighted means for baseline covariates of the ReCORD study are also presented.



\subsection{A simulation study}

To investigate the behavior of the proposed exact matching and the propensity score matching, we also conduct a simulation study in which 10,000 pairs of IPD's are randomly generated. In each simulation, one pair of IPD's is simulated, where each IPD has 300 patients and 15 covariates to be used for the matching. The 15 covariates are divided into 3 blocks of 5. Within each block, the five covariates followed a compound-symmetric structure (same correlation within block, but different for different blocks). Between any two of the 3 blocks, the covariates are independent. In addition, the 15 covariates are first simulated on a continuous scale; afterwards, 7 of them are dichotomized to binary variables, and 3 of them categorized to factors with more than 2 levels. The other 5 remain on the continuous scale.

The covariates in the two IPD's in each pair have identical covariance structure, but with the centers of the second IPD shifted. The same data structures are applied to all 10,000 pairs of simulated data sets. 

A normally distributed response variable $Y$ is simulated for both IPD's to depend on 6 of the 15 covariates. The 3 shifted variables are among the 6 influencers. The dependency is on the underlying continuous scales of these variables before they are dichotomized or categorized.

Details of the simulation process can be found in Appendix \ref{simproc}. As an example, the summary statistics of one of the simulated pairs are also given in Table \ref{tab:summ110} in Supplemental Material \ref{summ110}. The results of the exact matching and propensity score matching of the simulation study are discussed in Section \ref{simstudies}.

\section{Matching}\label{matching}

\subsection{Background: Matching by probability of selection}\label{propscore}

Consider a target population $T$ containing patients characterized by their covariates $\mathbf{x}_i, i=1, \ldots$. If we randomly select a patient from $T$, the probability of obtaining a patient with covariate $\mathbf{x}_i$ is $P_T(\mathbf{x}_i)=:q_i$. 

In another population $S$, patients have the same covariates as in $T$, but with a different distribution. If we randomly pick a patient from $S$, the probability of getting a patient with covariate $\mathbf{x}_i$ is $P_S(\mathbf{x}_i)=:p_i$.

In population $S$, we now modify the random selection procedure by assigning the weight $w_i=q_i/p_i$ to patient $i$. The probability of randomly selecting patient $i$ with covariate $\mathbf{x}_i$ from $S$ becomes $\frac{q_i}{p_i}p_i=q_i$. Hence, we have matched $S$ onto $T$. The weights $w_i$ are unscaled. If we are repeating the random selection $n_S$ times so to give us a sample of size $n_S$, and want to make sure that the weights add up to $n_S$, we have to scale them by $n_S\cdot \frac{w_i}{\sum_{i=1}^{n_S} w_i}$.


To introduce propensity score matching, assume that we observe data
$\left(z_i,\mathbf{x}_i,\mathbf{y}_i\right)$ from patients $i=1,\ldots, n$ in population $S$ where $z_i$ is an indicator of an intervention/exposure/treatment we are interested in, $\mathbf{x}_i$ is a vector of additional covariates and $\mathbf{y}_i$ is a vector of responses. 
$\left(\mathbf{x}_i,\mathbf{y}_i\right)$ are realizations of the random variables $(X,Y)$. For the propensity score matching, we interpret $z_i$ as the realisation of the random variable $Z$, and we call the quantity $P\left(Z=z_i\left|\mathbf{x}_i\right.\right)$ the \textit{propensity score}.



In this paper, we aim to compare two different treatments, each investigated in a different study, $0$ and $1$. Propensity score matching uses a logistic regression model treating the study indicator $z_i$ as the response and the patient characteristics $\mathbf{x}_i$ as covariates. The estimate $\hat{p_i}$ from this model is interpreted as an estimate of the propensity score $P\left(Z=1\left|\mathbf{x}_i\right.\right)=P_1(\mathbf{x}_i)$, the probability that a randomly selected patient with covariates $\mathbf{x}_i$ is in study $1$. $Z$ represents trial membership. As a consequence of this setup, $1-\hat{p}_i$ becomes the estimate of $P_0(\mathbf{x}_i)$. So if we want to match study $0$ onto study $1$, we would use the odds $\frac{p_i}{1-p_i}$ of being in study 1 as weights in study $0$. Conversely, we use the odds $\frac{1-p_i}{p_i}$ of being in study $0$ if we want to match study $1$ onto study $0$. These weights are again unstandardized. To retain the original sample size in study $0$, we would divide all weights by the sum of weights, then multiply with original sample size. In practice, however, it is more advisable to discount the sample size somewhat to adjust for the uncertainty in the weights. One idea is to have them sum up to the effective sample size (ESS, see below).

In the target population, we assign a weight of $1$. That is, if study $0$ is matched onto study $1$, the weights in study $1$ are all $p_i/p_i=1$. This is mostly a formality to facilitate comparison of responses after matching.

If we want to match all observations onto a pooled population from the two studies with the pool composed of fractions $\nu_k$ of patients from study $k=0,1$, $\nu_0+\nu_1=1$, then the weights are 
\begin{equation}\label{eq_wiprop}
w_i \propto
\left\{
\begin{array}{ll}
\frac{\hat{p}_i\cdot \nu_1+(1-\hat{p}_i)\cdot \nu_0}{1-\hat{p}_i} & \mbox{ for patient }\ i\ \mbox{in study } 0\\
\frac{\hat{p}_i\cdot \nu_1+(1-\hat{p}_i)\cdot \nu_0}{\hat{p}_i} & \mbox{ for patient }\ i\ \mbox{in study } 1
\end{array}
\right. .
\end{equation}
While in theory $\nu_k$ could be arbitrary numbers between $0$ 
and $1$, the two most natural choices are either to use the observed fractions $\nu_k=\frac{n_k}{n_0+n_1}$ or $\nu_k=0.5$. In the latter case, the weights are proportional to 
\begin{equation}\label{eq_weights}
w_i \propto
\left\{
\begin{array}{cl}
\frac{1}{1-\hat{p}_i} & \mbox{ for patient }\ i\ \mbox{in study } 0\\
\frac{1}{\hat{p}_i} & \mbox{ for patient }\ i\ \mbox{in study } 1
\end{array}
\right. .
\end{equation}
These coincide with the weights commonly used in epidemiology when considering a population where everyone is exposed. It is often called \textit{inverse probability of treatment weighting} (IPTW), see e.g. \citet{Austin2016}.

Finally, we note that the scaling of the weights is irrelevant to the comparison of point estimates of the treatment effect. Average responses are compared, and the division by the weights eliminates all scaling constants ($\frac{\sum_i w_i y_i}{\sum_i w_i}=\frac{\sum_i c\cdot w_i y_i}{\sum_i c \cdot w_i}$) from them. For variance estimation, however, this plays a role.

In this setup, the propensity score is viewed as a probability. In the example where $Z$ represents exposure and the interest is in assessing the impact of exposure on some outcome, the interpretation of the propensity score as the probability of being exposed (or not) given a person's characteristics seems natural. In examples, however, where $Z$ is a study index, interpreting the propensity score as a probability seems less straightforward to us. We prefer to regard it merely as a tool for matching populations.

Propensity score matching will typically lead to samples that are better matched. The standardized mean difference (SMD) is often used for an informal check of quality of the matching. For the unweighted data before matching, the SMD is defined as
$$
d_j = \frac{|\bar{x}_{j,1}-\bar{x}_{j,0}|}{s_{pooled}}
$$
where $\mathbf{x}_i=(x_{i1},\ldots,x_{ip})'$, $\bar{x}_{j,k}=\frac{\sum_{i:z_i=k} {x}_{ij}}{n_k}$, $s_{pooled}$ is the usual pooled standard deviation based on $s_{j,1}$ and $s_{j,0}$, $s_{j,k}$ being estimates of the standard deviation of $\left(x_{ij}\right)_{i:z_i=k}$ in study $k$, e.g. 
$$
s_{j,k}^2=\frac{1}{n_{k}-1}\sum_{i:z_i=k}\left(x_{ij}-\bar{x}_{j,k}\right)^2.
$$
For the matched data, the corresponding weighted versions are used. In particular, the weighted mean is defined as
$$
\bar{x}_{j,k}^*=\frac{\sum_{i:z_i=k}w_ix_{ij}}{\sum_{i:z_i=k}w_i},
$$
and the weighted estimated variance is
$$
{s_{j,k}^{*2}}=\frac{\sum_{i:z_i=k}w_i\left(x_{ij}-\bar{x}_{j,k}^* \right)^2}{\sum_{i:z_i=k}w_i}.
$$
Popular rules of thumb consider the two groups adequately matched if all SMDs $d_j$ are below 0.1 or 0.2 (after matching). There are no generally accepted guidelines regarding the measures to be taken when this requirement is not met after matching. Likewise, to our knowledge, there is no broad consensus on how the estimation of the standard deviation should be performed after matching. The version we propose above is only one among many options. Hence, the SMD can merely provide a rough ad hoc assessment of the quality of matching.

In special cases, propensity score matching can lead to perfect matching in the sense that $\sum_{i:z_i=1} w_i \mathbf{x}_i=\sum_{i: z_i=0} w_i \mathbf{x}_i$. This is possible if all covariates are categorical such that every study participant $i$ belongs to a unique category $c(i)$. For example, with the three variables ``baseline disease severity", ``age group" and ``sex", the category $c(i)$ could be the triplet (``mild", ``20-29 years", ``female"). 
In such situations, an exact match arises from fitting the saturated logistic regression model
$$
logit(p_i)=\mu_{c(i)}.
$$
Writing down the likelihood equations from the corresponding binomial distributions, it is readily verified that they are maximized by $\hat{p}_i=n_{1,c(i)}/n_{c(i)}$, i.e. the number of patients in category $c(i)$ which are in study $1$ divided by the number of all patients in category $c(i)$. Empty categories drop out. For categories $c(i)$ in which only one of the two studies is represented, the usual problem of quasi-complete separation arises, i.e. the ML-estimates do not fulfill the conditions for the usual asymptotics results such as Wilks' theorem in that case. In a model with three categorical covariates, say, the saturated model is equivalent to a model with intercept, all main effects, all two-way, and the three-way interactions. 
In general, however, weights derived from propensity scores will not lead to equal averages of the covariates in the two studies.

\subsection{Exact matching}\label{matching_exact}

As an alternative, we suggest to use exact matching by standardization. In this technique, individual weights are calculated in such a way that after matching, the population averages $\bar{\mathbf{x}}_0$ and $\bar{\mathbf{x}}_1$ are identical, not just similar.

For the sake of simplicity, let us assume that we want to compare two treatments which were applied in two different studies. The studies are deemed to be sufficiently similar in design, but not identical. We assume that in both studies individual patient data are available. Let $\mathbf{X}_k$, $k=0,1$ be the $n_k\times p$-matrix of $p$ covariates on every individual. The rows of $\mathbf{X}_k$ are the vectors $\mathbf{x}_{i}'$ containing the covariates of patient $i$ from study $k$. Let 
$$
\mathbf{X}={-\mathbf{X}_0 \choose  \mathbf{X}_1}'.
$$

We want to find a weighting $\mathbf{w}=\left(\mathbf{w}_0',\mathbf{w}_1'\right)'$ such that the averages of covariates in the first and the second study match perfectly, i.e. such that $\mathbf{Xw}=\mathbf{0}$ subject to $w_{ik}\geq 0$ for all $i,k$, $(\mathbf{1}_{n_0}',\mathbf{0}_{n_1}')\mathbf{w}=1$, $(\mathbf{0}_{n_0}',\mathbf{1}_{n_1}')\mathbf{w}=1$ (where $1$ is just a normalizing constant).

These are linear constraints, so linear programming, commonly known as LPsolve (with an arbitrary objective function) can be used to check whether a solution exists. \citet{glimm2022} have described this for the case of matching populations when individual patient data (IPD) are available from one data source, but only aggregate data (AD) from the targeted matching population. The technique can also be used when IPD are available on both sides. Since only linear programming is needed for it, the existence check is extremely quick, taking fractions of seconds on an ordinary PC, even with hundreds of variables. Usually, the mere check of existence of a solution is less interesting than in the IPD/AD case, because a match almost always exists in practice\footnote{A solution exists if the intersection of the two convex hulls of $\mathbf{X}_0$ and $\mathbf{X}_1$ in $p$-dimensional space is not empty, see \cite{glimm2022}.}, as will become clear when discussing the illustrative examples below.

A simple way to obtaining a unique solution would be to apply the matching-adjusted indirect
comparisons (MAIC) technique \citep{signorovitch2010} by assigning a weight of $1/n_k$ to everyone in study $k$, and then find MAIC weights (or other weights) for the other study. This would use the population from study $k$ as the target population.
A more versatile approach views weight matching as a constrained optimization problem. In addition to the linear constraints, this requires the selection of an appropriate objective function. While in general, all-purpose optimizers (such as \verb|optim()| in \verb|R|) can be used to find the weights, there are specialized optimizers for linear and quadratic programming that should be used if the objective function is linear or quadratic. In \verb|R|, \verb|LPSolve| for linear programming and \verb|quadprog| for quadratic programming perform the required optimizations very quickly, such that computation time is not an issue with the number of variables to be matched in typical applications (usually up to a few dozen, rarely hundreds, never thousands). The algorithm used by \verb|quadprog| solves the convex optimization problem in a finite number of steps (\citealp{goldfarb1983}, theorem 3).

One candidate objective function for our matching problem would be the aforementioned ESS, where
\begin{equation}\label{eq_ESS}
ESS\left(\left\{\mathbf{w}_k\right\}_{k=0,1}\right)=\frac{\left(\sum_i\sum_k w_{ik}\right)^2}{\sum_i\sum_k w_{ik}^2},
\end{equation}
which is equivalent to minimizing $\mathbf{w}'\mathbf{w}$. Note that this also maximizes\linebreak $\frac{(\sum_i\sum_k w_{ik})^2}{(\sum_i\sum_k w_{ik})^2-\sum_i\sum_k w_{ik}^2}$ which corresponds to a popular criterion in propensity score matching. Alternatively, imbalances in per-study-ESS might be penalized by using $\mathbf{w}'\mathbf{w}+\delta\left(\mathbf{w}_0'\mathbf{w}_0-\mathbf{w}_0'\mathbf{w}_1\right)^2$ as the objective function to minimize. Yet another criterion would be to maximize the sum of the two ESS's of each group. This is different from maximizing the entire ESS due to the definition of ESS as a ratio of squares.

The approach guarantees that after matching the weighted means of the covariates are the same in the two studies. However, there is no restriction that forces the overall weighted mean to be similar to the overall unweighted mean of the pool of the two studies. It can happen that the matched mean is not between the two means in one of the covariates (see section \ref{results_illu} for an example). To avoid this, we can add a corresponding requirement to the set of constraints:
    \begin{equation}\label{addconstraint}
    \left(\mathbf{X}_0', \mathbf{X}_1'\right)'\mathbf{w}\in\left[2\cdot\min\left(\bar{\mathbf{X}}_0,\bar{\mathbf{X}}_1\right), 2\cdot\max\left(\bar{\mathbf{X}}_0,\bar{\mathbf{X}}_1\right)\right],
    \end{equation}
    where $\bar{\mathbf{X}}_k$ is the vector of covariate averages in study $k$. This defines $p$ additional restrictions which are all of the form ``weighted average across studies is within the minimum and the maximum of unweighted averages per study". 

Additional constraints could be added to limit the maximum size of single weights $w_{ik}$. Obviously, any additional constraints could render an otherwise solvable problem unsolvable. 
All of the mentioned suggestions lead to a quadratic programming problem for which the package \verb|quadprog| is available in \verb|R|. In the examples of section \ref{results}, we minimized $\mathbf{w}'\mathbf{w}$. These calculations can be performed using the R package \verb|maicChecks| (version 0.2.0). Appendix \ref{apx_ExMa} provides more details on the quadratic programming algorithm.

Finally, we might relax the equality $\mathbf{Xw}=\mathbf{0}$ to $\mathbf{Xw}\in \left[-\bm \epsilon,\bm \epsilon\right]$ or to $\mathbf{w'X'AXw}\leq \epsilon$ if we are willing to accept non-zero differences between the means after matching. $\mathbf{A}$ would typically be a matrix related to the empirical variance of the variables in $\mathbf{X}_0$ and $\mathbf{X}_1$, e.g. the inverse of the pooled covariance estimate or the diagonal matrix with the inverse of the pooled variance estimates on the diagonal. The difficulty with this is to find appropriate thresholds $\bm \epsilon$ or $\epsilon$. For this reason, we will not pursue this approach further here.  

\subsection{More on comparisons with other methods}

We have already contrasted our approach with propensity score matching in section \ref{propscore}, but there are some other closely related approaches which we might also want to compare with.

MAIC \citep{signorovitch2010,signorovitch2012} is an approach for matching individual-patient data (IPD) onto aggregated data data (AD). Like the approach suggested here, it enforces perfect matching of the weighted IPD means onto the reported AD means. Like propensity score matching, but unlike the suggestion here, it restricts the weights to have a specific functional form, namely $w_i\propto \exp(\mathbf{y}_i'\bm \beta)$. Similarly, \citet{alsop2022} suggest to restrict weights to have a polynomial form. Since the exponential function can be closely approximated by a polynomial, these approaches give very similar results. \citet{alsop2022} also mention the possibility of ``fuzzy" matching by introducing tolerance limits for the difference between weighted IPD averages and the reported AD averages.

\section{Results of the exact matching}\label{results}

In this section, we illustrate the application of the suggested matching approach with the examples introduced in section \ref{motivation}. Throughout this section, formula (\ref{eq_ESS}) is used as the objective function for calculating the exact matching weighted means.

\subsection{Illustrative simulated data}\label{results_illu}
\begin{figure}
    \centerline{\includegraphics[width=1.2\textwidth]{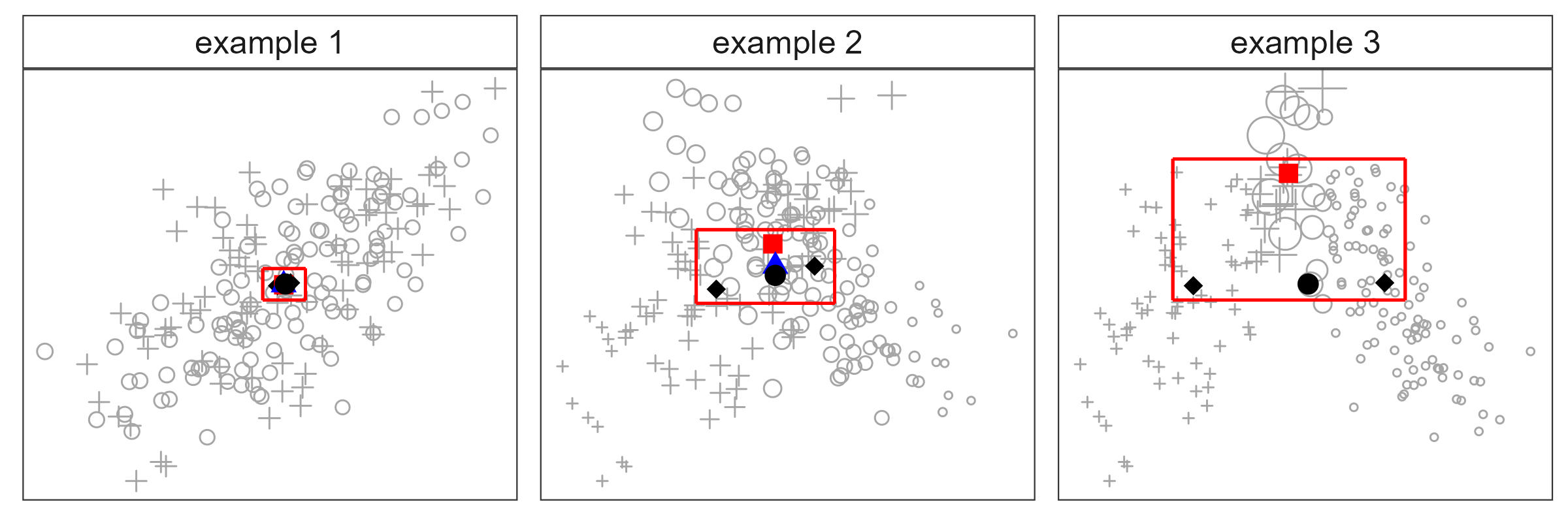}}
    \centerline{\includegraphics[width=1.2\textwidth]{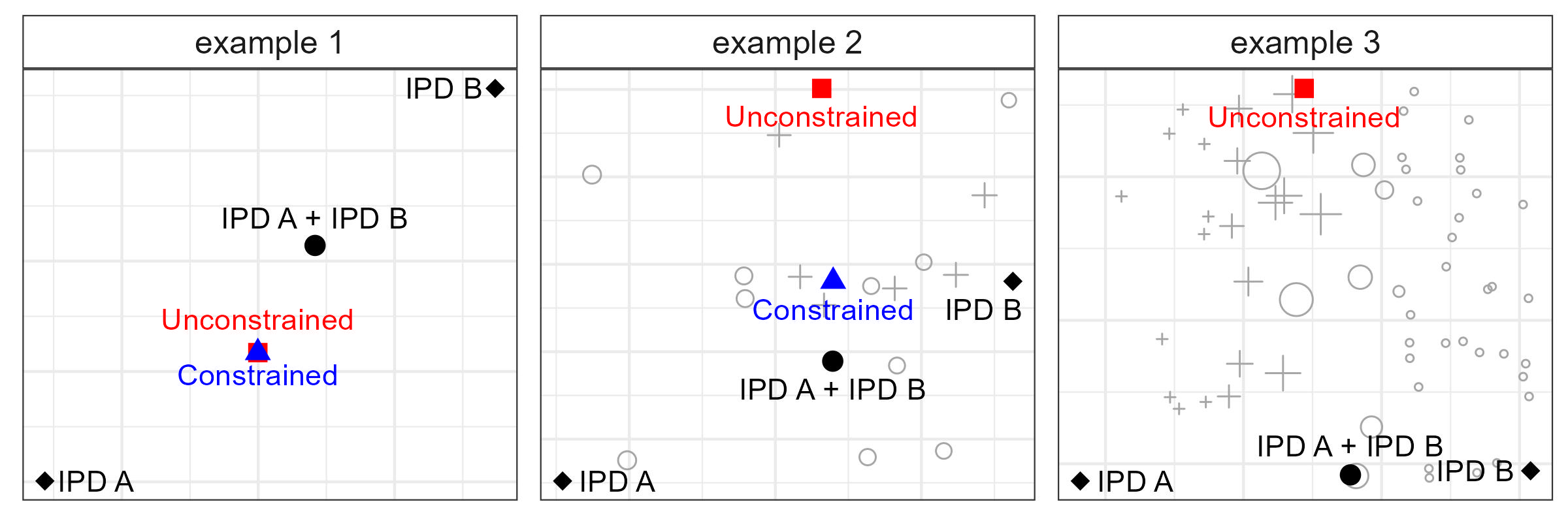}}
    \centering \footnotesize{Upper panel: weighted mean ({\color{red}{$\smallblacksquare$}}) and weighted mean with additional constraint 
 ({\color{blue}{$\smallblacktriangleup$}}) in relation to the individual weighted observations from IPD A (+) and IPD B ($\medcircle$), the observed (unweighted) IPD means ($\blackdiamond$), and the observed overall means of IPD A and IPD B combined (\textbullet).\\ Lower panel: details in the red boxes in the upper panel.\\ $x_1$: horizontal axis; $x_2$: vertical axis} 
    \caption{Weighted means of the simulated examples}
    \label{fig:plt2}
\end{figure}  

Figure \ref{fig:plt2} illustrates weights that each observation receives using the exact matching procedure from section \ref{matching_exact} for the simulated examples from section \ref{sub_illuexp}. The upper panel are the scatter plots as in Figure \ref{fig:plt0}, only the symbols for the observations are proportional to the weight they receive: the higher the weight the larger the symbol. The means of $x_1$ and $x_2$, both observed and weighted, are super-imposed on the scatters. They are also produced separately as a zoomed version in the lower panel of Figure \ref{fig:plt2}. 

For the ``well mixed" example 1, the raw means of the covariates in IPD A and IPD B and the constrained and unconstrained means of the covariates after matching are very close together. Here, the unconstrained mean already fulfills the additional constraints (\ref{addconstraint}), so the constrained and unconstrained means are identical.

For example 2, there is sufficient overlap to calculate both a constrained and an unconstrained mean, but they are no longer the same. The second variable ($x_2$ along the vertical axis) in the unconstrained mean has a higher value than the two corresponding values from the IPD sets. If the constraints are used, this value is shifted down to be identical with the average of $x_2$ in IPD B in this case, and is inside the two observed averages of IPD A and IPD B. This is the purpose of the additional constraints.

For example 3, the overlap between IPD A and IPD B is still sufficient to obtain an unconstrained common exact matched mean, but they are too different for exact matching with the additional constraint. This is an example where additional constraints render a solutiom impossible. The reason here is that within the rectangle defined by the observed IPD A mean and the observed IPD B mean, the two datasets do not intersect.


\subsection{Matching of data from the trials ReCORD and ELARA}\label{results_real}

Table \ref{tab:ReEl} shows the summary statistics of the baseline covariates for studies ReCORD and ELARA \citep{hampson2024}. Also presented are the weighted means of the constrained and the unconstrained exact matching as introduced in Section \ref{matching_exact}. In addition, we provide the propensity score weighted means for each covariate calculated from the propensity score matching using Equation (\ref{eq_wiprop}). Note that these weights differ from the propensity odds weights presented in \cite{hampson2024}.

As discussed in the previous sections, the exact matching weighted means in the two studies are identical, whereas there are still differences between weighted means when propensity score matching is used. Consequently, the SMDs for all exact matching weighted means are 0 and only the SMDs for the propensity score weighted means are shown in Table \ref{tab:ReEl}.  For three of the covariates, the unconstrained exact matching weighted means do not fulfill condition (\ref{addconstraint}). The unconstrained means of these covariates are in boldface in Table \ref{tab:ReEl}. The constrained exact matching means hence are moved onto the closer of the two observed means. For variables where condition (\ref{addconstraint}) is not violated, constrained and unconstrained means can differ, but in this example, these differences are minimal.

The ESS of the exact matched weights is higher than that of the propensity score weights for ELARA, and lower for ReCORD. Two phenomena are at play here: Firstly, propensity scores are not subject to the same restrictions as exact matching weights, so they can have a higher ESS than even the best weights obeying these restrictions. Secondly, in general the variance of the difference between independent means from two random samples with equal variance per sample unit is minimized when the sample sizes are equal. Because of this, maximizing the ESS of formula (\ref{eq_ESS}) induces a tendency to downweigh the larger sample (ReCORD) more than the smaller sample (ELARA).

\begin{figure}
    \centering
    \includegraphics[width=1\textwidth]{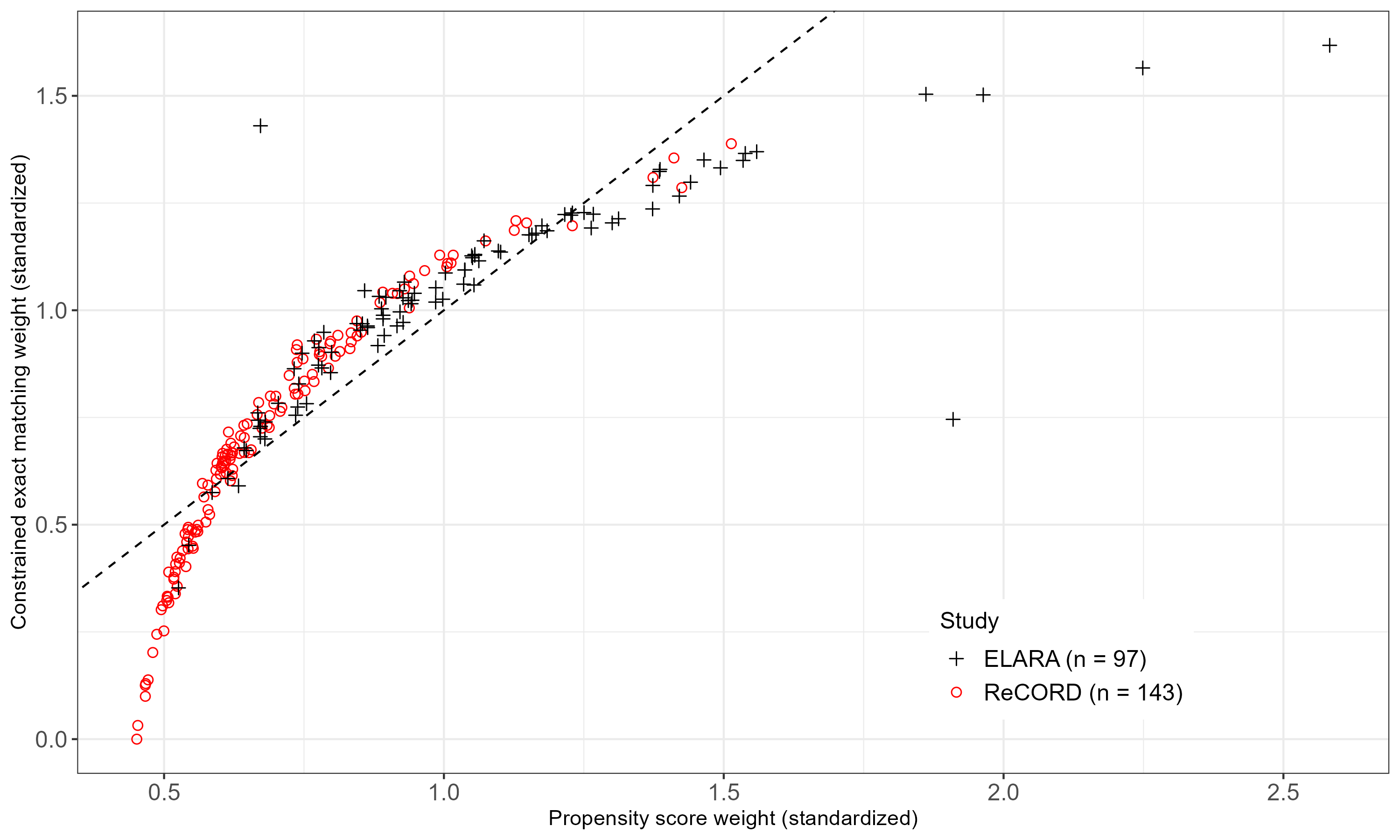}   
    \caption{Standardized propensity score weights vs. standardized constrained exact matching weights}
    \label{fig:plt7}
\end{figure}

\begin{figure}[ht]
    \centering
    \includegraphics[width=1\textwidth]{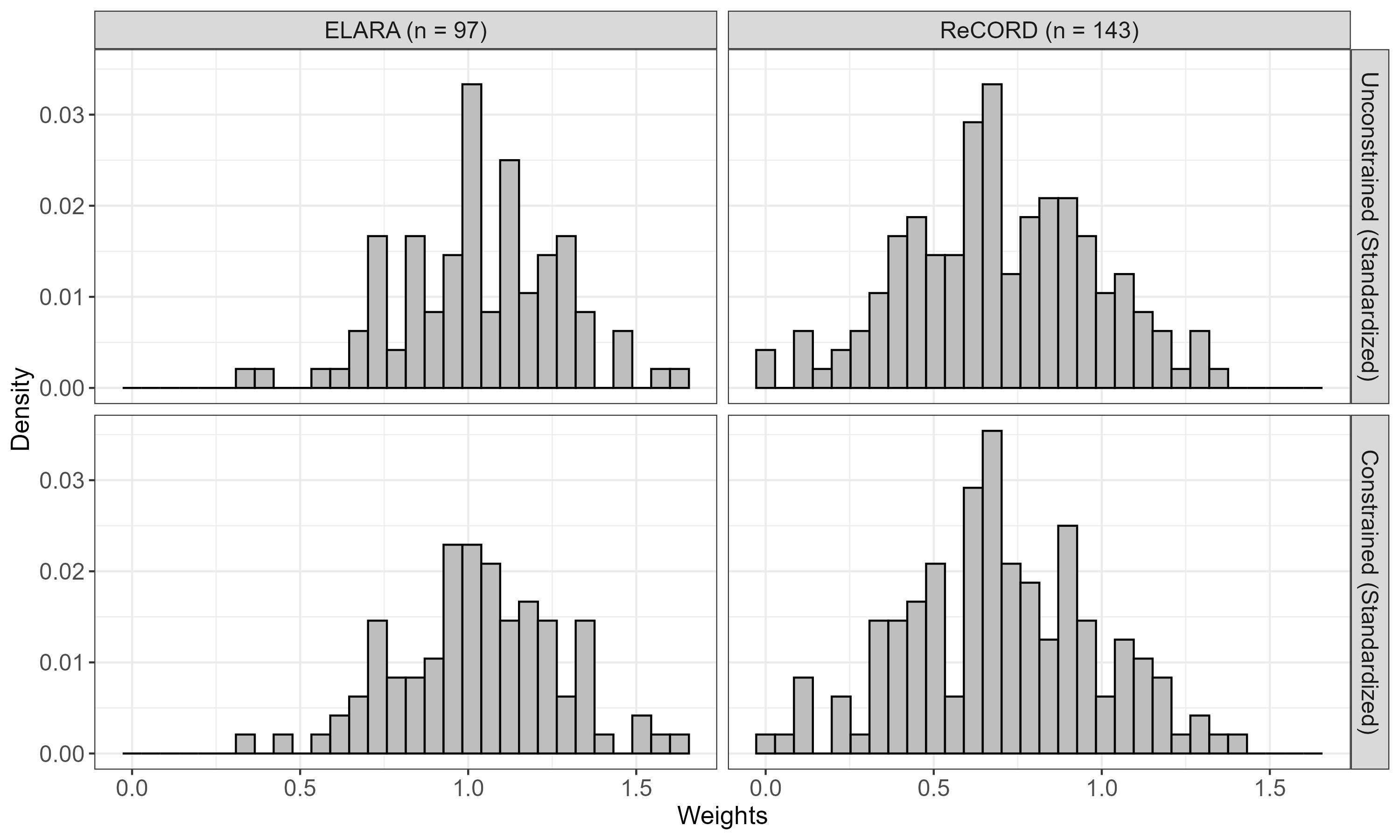}
    \caption{Histograms of unconstrained and constrained exact matching weights for ELARA vs ReCORD}
    \label{fig:histReEl}
\end{figure}

Figure \ref{fig:plt7} plots the propensity score weights versus the constrained exact matching weights. To render them comparable, both sets of weights have been standardized to add up to $100$ per study. Hence, the ELARA weights are on average larger than the ReCORD weights. The shape of the curve reflects the use of a quadratic objective function in contrast to an exponential weight function. The constrained exact matching weights are ``less extreme" in the sense that there are less individuals getting very high weights. They can drop to 0, however, whereas this cannot happen with propensity score weights. This can also be seen in the histograms of Figure \ref{fig:histReEl}. 

The matching behaves well in this example since the studies are quite similar in the covariates even before matching. Thus, the use of either propensity score matching or exact matching leads to the same conclusions and the choice of method becomes a matter of personal preference.

\newpage
\begin{sidewaystable}[!ht]
\footnotesize
    \centering
    \begin{tabular}{lccccccccccc} \hline
& \multicolumn{2}{c}{} & & \multicolumn{4}{c}{Exact matching weighted means} & & \multicolumn{3}{c}{} \\ 
    & \multicolumn{2}{c}{Observed means} & & \multicolumn{2}{c}{Unconstrained} & \multicolumn{2}{c}{Constrained} & &\multicolumn{3}{c}{Propensity score weighted means} \\ \cline{2-3} \cline{5-8} \cline{10-12} 
    & ELARA & ReCORD & & ELARA & ReCORD & ELARA & ReCORD & & ELARA & ReCORD & $|$SMD$|$ \\ \hline
Sample size & $97$ & $143$ & & & & & & & & & \\
ESS & & & & 92.0 & 122.8 & 92.0 & 122.5 & & 86.5 & 131.7 & \\
\multicolumn{12}{l}{Age (years) at treatment initiation (mean)} \\
& 56.5 & 60.1 & & \multicolumn{2}{c}{57.5} & \multicolumn{2}{c}{57.6} & & 58.0 & 58.5 & 0.040 \\ 
\multicolumn{12}{l}{Region (\%)} \\
$\quad$ Europe & 45.4 & 62.9 & & \multicolumn{2}{c}{50.7} & \multicolumn{2}{c}{50.2} & & 52.9 & 54.3 & 0.029 \\
$\quad$ Rest of World & 54.6 & 37.1 & & \multicolumn{2}{c}{49.3} & \multicolumn{2}{c}{49.8} & & 47.1 & 45.7 & 0.029\\ 
\multicolumn{12}{l}{Gender (\%)} \\
$\quad$ Male & 66.0 & 57.3 & & \multicolumn{2}{c}{64.6} & \multicolumn{2}{c}{64.7} & & 63.9 & 62.1 & 0.037\\
$\quad$ Female & 34.0 & 43.7 & & \multicolumn{2}{c}{35.4} & \multicolumn{2}{c}{35.3} & & 36.1 & 37.9 & 0.037 \\
\multicolumn{12}{l}{Prior autologous HSTC (\%)} \\
{$\quad$ Yes} & 37.1 & 37.1 & & \multicolumn{2}{c}{\bf 38.3} & \multicolumn{2}{c}{37.1} & & 38.3 & 36.8 & 0.030\\
$\quad$ No & 62.9 & 62.9 & & \multicolumn{2}{c}{\bf 61.7} & \multicolumn{2}{c}{62.9} & & 61.7 & 63.2 & 0.030\\
\multicolumn{12}{l}{Number of previous lines of systemic treatment (\%)} \\
$\quad\ 2-4$ & 71.1 & 76.9 & & \multicolumn{2}{c}{72.9} & \multicolumn{2}{c}{73.1} & & 74.8 & 74.3 & 0.013 \\
$\quad\ >4$ & 28.9 & 23.1 & & \multicolumn{2}{c}{27.1} & \multicolumn{2}{c}{26.9} & & 26.2 & 25.7 & 0.013\\
\multicolumn{12}{l}{Months between diagnosis and initiation of treatment (Mean)} \\ 
& 77.3 & 72.1 & & \multicolumn{2}{c}{75.4} & \multicolumn{2}{c}{75.3} & & 74.9 & 74.1 & 0.016 \\
\multicolumn{12}{l}{Failing to respond within 24 months to first-line therapy (\%)} \\ 
$\quad$ Yes & 62.9 & 60.1 & & \multicolumn{2}{c}{61.9} & \multicolumn{2}{c}{61.9} & & 61.6 & 61.4 & 0.003\\
$\quad$ No & 37.1 & 39.9 & & \multicolumn{2}{c}{38.1} & \multicolumn{2}{c}{38.1} & & 38.4 & 38.6 & 0.003\\
\multicolumn{12}{l}{Double refractory (\%)} \\
$\quad$ Yes & 68.0 & 67.8 & & \multicolumn{2}{c}{\bf 68.7} & \multicolumn{2}{c}{68.0} & & 68.1 & 68.1 & 0.000\\
$\quad$ No & 32.0 & 32.2 & & \multicolumn{2}{c}{\bf 31.3} & \multicolumn{2}{c}{32.0} & & 31.9 & 31.9 & 0.000\\
\multicolumn{12}{l}{Number of nodal involvement at treatment initiation (\%)} \\
$\quad\ \le 4$ & 59.8 & 48.3 & & \multicolumn{2}{c}{56.5} & \multicolumn{2}{c}{56.2} & & 55.2 & 53.7 & 0.031\\ 
$\quad\ >4$ & 40.2 & 51.7 & & \multicolumn{2}{c}{43.5} & \multicolumn{2}{c}{43.8} & & 44.8 & 46.3 & 0.031 \\
\multicolumn{12}{l}{Disease stage at initial diagnosis (\%)} \\
$\quad$ 1  & 6.2 & 7.0 & & \multicolumn{2}{c}{5.8} & \multicolumn{2}{c}{\bf 6.2} & & 5.4 & 6.1 & 0.027\\
$\quad$ 2 & 13.4 & 9.1 & & \multicolumn{2}{c}{10.1} & \multicolumn{2}{c}{10.1} & & 9.6 & 9.3 & 0.010\\
$\quad$ 3 & 21.6 & 18.2 & & \multicolumn{2}{c}{21.5} & \multicolumn{2}{c}{21.5} & & 20.7 & 21.2 & 0.014\\
$\quad$ 4 & 58.8 & 65.7 & & \multicolumn{2}{c}{62.6} & \multicolumn{2}{c}{62.3} & & 64.3 & 63.4 & 0.020\\ \hline
    \end{tabular}
    \caption{Observed means, exact matching weighted means (unconstrained and constrained), and propensity score means with standardized mean difference (SMD)}
    \label{tab:ReEl}
\end{sidewaystable}

\clearpage

\subsection{Simulation studies}\label{simstudies}

Based on the matching of the 10,000 simulated pairs of IPD's, we see that between the two methods, exact matching, either the unconstrained or the constrained version, offers a more consistent result. This can be seen in the distributions of the ESS that each IPD receives after matching (Figure \ref{fig:histESS_sim} below and Table \ref{tab:summess} in Supplemental Material). For exact matching, either unconstrained or constrained, the ESS is centered around 140 for IPD A and 150 for IPD B, with a range of 95 to 191. The standard deviations of the 10,000 pairs of ESS is between 10 and 11 for both IPD's and for both versions of the exact matching. 

In contrast, using the propensity score weights, the ESS vary significantly from data set to data set. On the one end, the ESS can be as low as 1.1, whereas on the upper end, the maximum is as high as 208. The standard deviations for the two IPD's are roughly 39 and 44, respectively (cf. Table \ref{tab:summess}).

\begin{figure}
    \centering
    \includegraphics[width=1\textwidth]{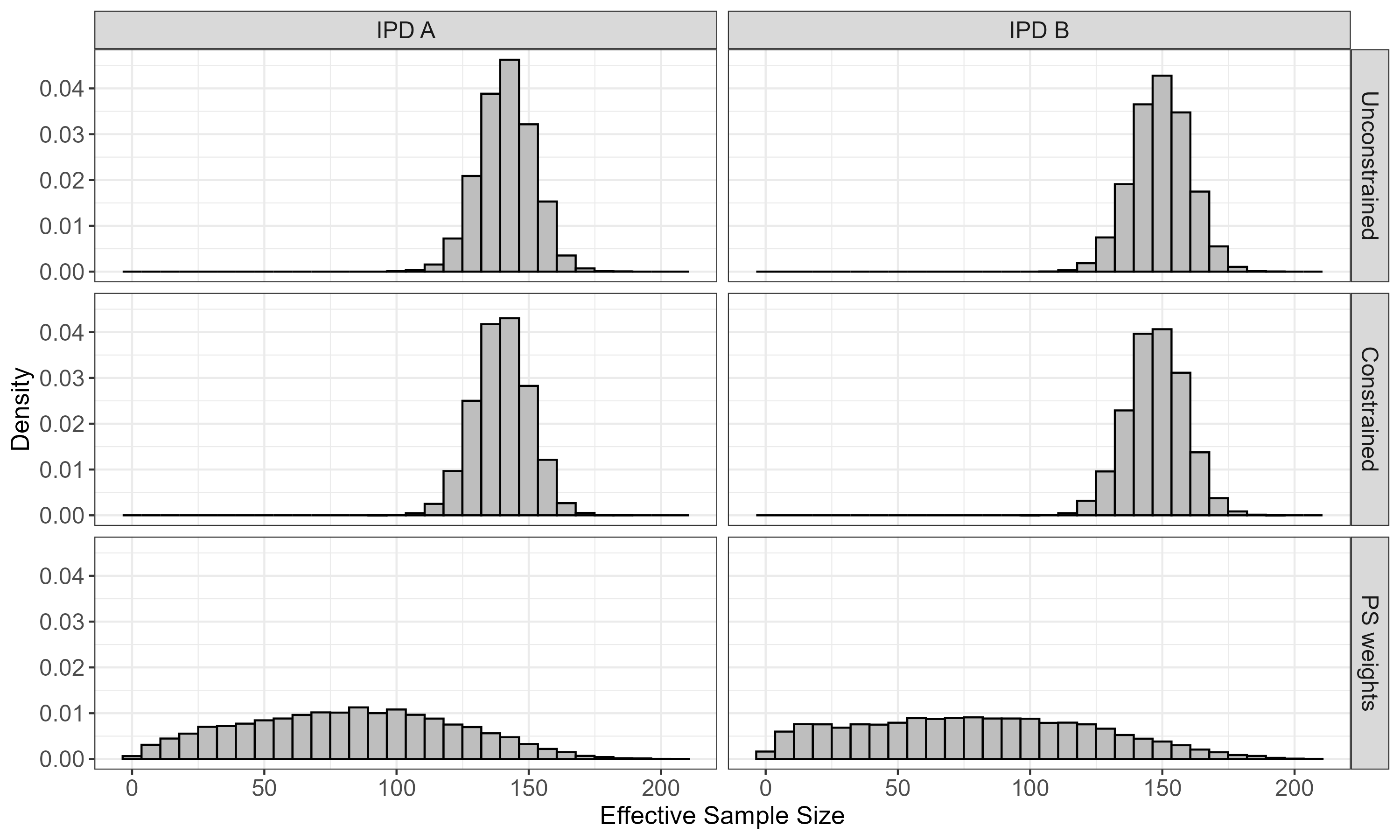}
    \caption{Histograms of ESS of matching 10,000 pairs of simulated IPD's using the exact method (unconstrained and constrained) and propensity score method}
    \label{fig:histESS_sim}
\end{figure}

The ESS is inversely related to the variation in the weights assigned to each observation in a data set, which in turn is dominated by the largest weight (since the smallest is bounded below by 0). Therefore, it is instructive to see the range of the 10,000 largest standardized weights that each matching produces (Table \ref{tab:maxwts}). 
Exact matching leads to a largest weight in the range of 1.1 to 3.2 for both unconstrained and constrained matching. For the propensity score weights, however, the largest weights can be anywhere from 2.5 to 94.0. The phenomenon of very large weights is a common issue in propensity score matching. As an ad hoc solution to this, restrictions on the highest weight are often imposed \citep{lee2011}. 

\begin{table}[ht]
\centering
\begin{tabular}{lrrr}
  \hline
Statistics & Unconstrained & Constrained & Propensity score  \\ 
& IPD A \& IPD B & IPD A \& IPD B & IPD A \& IPD B \\ \hline
\# of simulations & 10000 & 10000 & 10000 \\ 
Minimum    & 1.145   & 1.156   &  2.533   \\ 
1st quartile & 1.524   & 1.551   &  6.877   \\ 
Median  & 1.644   & 1.674   &  9.951   \\ 
Mean    & 1.668   & 1.700   & 13.239   \\ 
3rd quartile & 1.784   & 1.820   & 15.686   \\ 
Maximum    & 3.133   & 3.218   & 94.046   \\ \hline
Standard deviation & 0.206 & 0.213 & 10.383 \\ 
NA's    & 0 & 46   & 0 \\ \hline
\end{tabular}
\caption{Summary statistics of the largest standardized weights that each pair of IPD's received from each matching method among the 10,000 simulations} 
\label{tab:maxwts}
\end{table}

As briefly discussed in Section \ref{matching_exact}, weight calculation may fail if the two studies are too different. In theory, this can affect all three methods. In this simulation study, constrained exact matching had no solution for 46 of the 10,000 simulations. For unconstrained exact matching and for propensity score matching, there was no case without a solution. 

\begin{figure}
    \centering
    \includegraphics[width=1\textwidth]{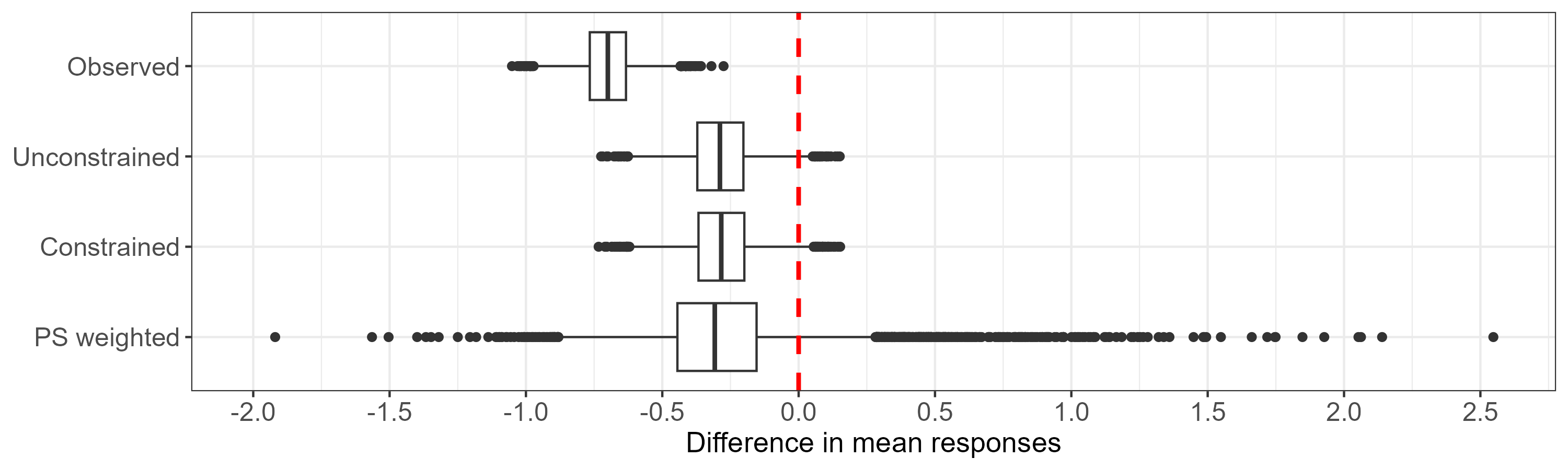}
    \caption{Boxplots of the mean difference in the response variable $Y$ among the 10,000 pairs of IPD's. $Y$ depends on the underlying continuous scale of 6 covariates.}
    \label{fig:boxYbar_num}
\end{figure}

Regarding the response variable $Y$, all three methods reduce the bias (in absolute value) of the estimated mean difference between the two studies considerably from unweighted 0.7 to weighted 0.3 (see Figure \ref{fig:boxYbar_num}). However, the bias is not completely eliminated. The reason is that the dependency of $Y$ is on the underlying continuous scale of the covariates, whereas the matching is done on the corresponding dichotomized or categorized version. When the response variable $Y$ is simulated to depend on the dichotomized or categorized covariates directly, the bias is eliminated by all three methods (see Figure \ref{fig:boxYbar_cat} in Supplemental Material \ref{suppsim}).
As a consequence of the large variability in the propensity score weights, the weighted differences estimated with propensity scores are generally much less stable than the difference estimates with exact matching. This is also seen in Figure \ref{fig:boxYbar_num}.

\section{Statistical considerations of estimating response to treatment}\label{statcon}

While the focus of this paper is on matching patient populations, the ultimate aim of all indirect comparison procedures is to compare treatments with respect to a response variable. A comprehensive overview of this topic \citep[see e.g.][]{daniel2018} is beyond the scope of this paper. However, some remarks are in order.

The average response in the matched population (i.e. the population onto which we match our data) is estimated from study $k$ as
$$
\hat{\bm \mu}_k=\frac{\sum_{i:z_i=k} w_i \bm y_i}{\sum_{i:z_i=k} w_i}.
$$
Under the assumption that $E(Y|X,k)=E(Y|X)$ for all studies $k$, this is a consistent estimate of 
$E_m(Y)$ where $m$ denotes the matched population. Note that if $z_i$ is not used for matching, but is different in $k=0,1$, then $E(Y|X,k)=E(Y|X)$ pertains to a null hypothesis of no differences between the two treatments $Z=0,1$.

Regarding the (co-)variance of $\hat{\bm \mu}_k$, some further definitions are required. For simplicity, assume that $Y$ is univariate. 
Modelling the distributions of $Y|X$ and of $X|k$ explicitly is difficult, especially if $X$ is multidimensional. If we are willing to accept the notion of 
$(z_i,\mathbf{x}_i,y_i), i=1,\ldots, n_k$ as an i.i.d. sample from the joint distribution $\left(Z,X,Y\right)$ in study $k$, we might use the bootstrap or a jackknife estimator to approximate the corresponding variances. 

Alternatively, we can condition on the observed values of $\mathbf{x}_i, i=1,\ldots,n_k$ in study $k$  and assume that $(y_i|z_i,\mathbf{x}_i)$ are stochastically independent. Thus,
since $\mathbf{x}_i$ uniquely determine $w_i$, we have
$$
var(\hat{\mu}_k|\mathbf{X}^{(k)})=\frac{\sum_{i:z_i=k} w_i^2 var(y_i|\mathbf{x}_i)}{\left(\sum_{i:z_i=k} w_i\right)^2}
$$
with $\mathbf{X}^{(k)}$ denoting the collection of $\mathbf{x}_i$ observed in study $k$.
We make the following assumptions:
\begin{enumerate}
\item\label{ass1} $var(y_i|\mathbf{x}_i, k)=var(y_i|\mathbf{x}_i)$ for all studies $k$,
\item\label{ass2} $var(y_i|k)$ is the same for all individuals $i$ within study $k$,
\item\label{ass3} $var(y_i)\geq var(y_i|\mathbf{x}_i)$ for all $\mathbf{x}_i$.
\end{enumerate}
The first two assumptions are common causal inference assumptions: The first assumption is the well-known ``no hidden confounders"-assumption. The second assumption is fulfilled if within study, the sample units $(\mathbf{x}_i,y_i)$ are exchangeable. Loosely speaking, this is  a minor relaxation of the assumption that $(\mathbf{x}_i,y_i)_{i\in k}$ is a random sample from study population $k$. The third assumption is a sufficient requirement made for convenience. 

Under these assumptions, we can estimate $var(\hat{\mu}_k|\mathbf{X}^{(k)})$ by
$$
\hat{var}(\hat{\mu}_k|\mathbf{X}^{(k)})=\frac{\sum_{i:z_i=k} w_i^2}{\left(\sum_{i:z_i=k} w_i\right)^2} s_k^2
$$
with $s_k^2=\frac{1}{n_k}\sum_{i:z_i=k} \left(y_i-\bar{y}_k\right)^2$, since $s_k^2$ is a consistent estimate of $var(Y|k)$ where $Y$ is a single random pick from the distribution of responses $Y$ in study population $k$.
Assumption \ref{ass3} guarantees that $\hat{var}(\hat{\mu}_k|\mathbf{X}^{(k)})$ is a conservative estimate of $var(\hat{\mu}_k|\mathbf{X}^{(k)})$. The assumption may be viewed as plausible because by the law of total variation $$
var(Y)=var(E(Y|X))+E(var(Y|X)),
$$
so $var(Y)>E(var(Y|X))$ unless $var(E(Y|X))=0$ which can only happen if $X$ deterministically predicts $Y$. In ordinary regression, a common assumption is that the variance of $Y$ does not depend on $X$, even if $E(Y|X)\neq E(Y)$. This is more restrictive than assumption \ref{ass3}. Otherwise, the condition implicitly requires that $var(Y|X)$ must not be ``too different" for different $X$.

Note that $\frac{\sum_{i:z_i=k} w_i^2}{\left(\sum_{i:z_ik} w_i\right)^2}=ESS^{-1}$ and that
$\hat{var}(\hat{\mu}_k|\mathbf{X}^{(k)})\geq \frac{1}{n_k}s_k^2$ with equality only if all $w_i \propto 1$. From this perspective, the $ESS$ can be viewed as a measure of the precision of the estimate. This also implies that matching can cost precision, e.g., if a covariate has a different distribution in the two studies but must be matched.  We would like to emphasize, however, that improving precision will in general lead to a loss of accuracy, e.g. by allowing differences between the average covariate values to persist post-matching as in propensity score matching or as a consequence of leaving covariates out of the matching. 

The calculated quantities $\hat{\mu}_k$ and $\hat{var}(\hat{\mu}_k|\mathbf{X}^{(k)})$ can subsequently be used for comparing the responses in the two studies, e.g. for testing treatment efficacy, the construction of confidence intervals and so on.

\quad



\section{Discussion}\label{discussion}

In this paper, we suggest to use exact matching via standardization to render the observations from two IPD sets comparable. 



On a technical level, the suggested method has much in common with propensity score matching. On the conceptual level, however, there are important differences. 

The  method we have introduced creates a common population from the two IPD sets. Technically, we \textit{could} also declare one of the two IPD sets as the target. In that case, every patient in that set would get a weight of 1 and only the other IPD set would be matched. This approach is appropriate if data from an RCT is compared with external data from a larger database. However, we are considering applications where none of the two IPD sets is a reasonable representative of the target population. We could assign one of them to be, but that choice would be arbitrary. The conclusions can differ depending on which one we pick as the target \citep{phillippo2018}. When comparing two RCTs that both represent the target population to some extent, they both should be reweighted.

Our approach does not attempt a formal definition of the target population. We
believe that in this aspect, our approach is close to the practice of clinical trials:
The inclusion and exclusion criteria of clinical trials, the choice of participating centers and other trial aspects invariably have a strong pragmatic element to them. For example, pregnant women, children and other vulnerable patient groups are excluded from many studies for safety and ethical reasons, not because they do not need treatment. Moreover, having a more homogeneous study population provides a more precise estimation of treatment effect. Hence, practitioners of clinical trials acknowledge the fact that the trial results from a moderately restricted patient population can still be generalized to a wider population in clinical practice.


While we acknowledge that this implicitness regarding the target population can be viewed as unsatisfactory, we believe that the solution offered to this problem by propensity score matching is even less satisfactory. Propensity score matching assumes the existence of a super-population for which the pooled data is representative. It equates ``trial participation" with ``exposure to treatment". However, none of the patients who participated in ReCORD or ELARA was offered to participate in the other study. 
Hence, the propensity, i.e. the ``probability of participating in ReCORD rather than ELARA", in our view lacks credibility as a probability measure. In addition, it is not infrequent that propensity score matching produces extremely large weights for some patients, as in the simulations of Section \ref{simstudies}. A common solution is to post-hoc truncate weights at some threshold. This further weakens the probabilistic interpretation of propensity scores.

With this interpretation gone, we are left with the disadvantages of non-exact matching. In that respect, we believe that the suggested exact matching method  is a better choice. The weights each patient receives are simply numerical scores that render the studies comparable without further probabilistic interpretation. 

Variance estimation of the weighted response discussed in Section \ref{statcon} is a consequence of this interpretation of the exact matching weight. 
While for decades, trialists have applied conditional statistical inference on the observed values of covariates within a given trial, this practice has recently drawn some criticism. This criticism is certainly not completely unwarranted. The additional uncertainty from the variation in covariates is ignored when conditioning on them. Still, this practice is arguably more defensible in randomized clinical trials than in observational studies or other non-randomized comparisons. Conditional tests of treatment effects which keep type I error under every condition also keep them unconditionally. Likewise, conditional confidence intervals have the right coverage for any arbitrary observed constellation of covariates. In most fields of statistics beyond clinical trials, conditioning on nuisance parameters is a widely accepted practice.

In conclusion, for comparisons of different treatment effects, running a clinical trial is still the gold standard. When this is not feasible, propensity score matching is the current default approach to indirect comparisons. The method, however, has been initially designed for epidemiological investigations where randomized studies are truly impossible. In this paper we describe an alternative method that is more targeted to indirectly comparing clinical trials. 

\bibliography{references}


\appendix
\section{Appendix}\label{appendix}\label{apx}

\subsection{Setup of the constrained optimization problem using the R function quadprog}\label{apx_ExMa}

The exact matching method described in section \ref{matching_exact} is implemented in the update R package \verb|maicChecks|. This is an outline of the setup of the convex quadratic optimization problem that \verb|maicChecks| uses internally for the calculation. 

\verb|R quadprog| solves quadratic optimization problems of the type
$$
\mathbf{b}'\mathbf{w}+\mathbf{w}'\mathbf{Q}\mathbf{w}
$$
subject to linear constraints $\mathbf{A}'\mathbf{w}\geq \mathbf{c}$ or $\mathbf{A}'\mathbf{w}=\mathbf{c}$.
In this case, the objective function is the ESS, that is, $\mathbf{Q}=\mathbf{I}_n$ and $\mathbf{b}=\mathbf{0}_n$ where $n_j$ is the number of sample units in study $j=0,1$ and $n=n_0+n_1$.
Regarding the constraints, we have the following setup:
\begin{eqnarray*}
{-\bm X_0 \choose \bm X_1}'\cdot \bm w=\bm 0_p\\
\left(\bm 1_{n_0}' \bm 0_{n_1}'\right)\cdot \bm w=1\\
\left(\bm 0_{n_0}' \bm 1_{n_1}'\right)\cdot \bm w=1\\
\bm I_n\cdot \bm w\geq 0,\\
\end{eqnarray*}
such that
$$
\bm A=\left(\begin{array}{ccccc}
-\bm X_0 & \bm 1_{n_0} & \bm 0_{n_0} & \bm I_{n_0} & \bm 0_{n_0}\\
\bm X_1 & \bm 0_{n_1} & \bm 1_{n_1} & \bm 0_{n_1} & \bm I_{n_1}
\end{array}\right)
$$
and $\mathbf{c}=\left(\mathbf{0}_p,1,1,\mathbf{0}_n\right)'$.
This setup guarantees that the weighted averages in the two groups are identical, $\mathbf{X}_1 \mathbf{w}_1=\mathbf{X}_2\mathbf{w}_2$ (if the constraints are feasible with the data), but there is no restriction on the overall weighted mean. In order to force the weighted mean to be between the two study means, we can add two additional constraints:
\begin{eqnarray}
{\bm X_0 \choose \bm X_1}'\cdot \bm w \geq 2\cdot\min\left(\bar{\mathbf{X}}_0,\bar{\mathbf{X}}_1\right) \label{A1}\\
-{\bm X_0 \choose \bm X_1}'\cdot \bm w \geq -2\cdot\max\left(\bar{\mathbf{\mathbf{X}}}_0,\bar{\mathbf{X}}_1\right), \label{A2}
\end{eqnarray}
where the minimum and the maximum are taken per row.
Matrix $\mathbf{A}$ and vector $\mathbf{c}$ must be augmented correspondingly. The factor 2 arises because of the constraints $\mathbf{1}_{n_j}'\mathbf{w}_j=1$ which means that $\mathbf{1}_n' \mathbf{w}=2$.

An additional issue arises if $(k-1)$ columns of $\mathbf{X}$ represent a dummy coding of a categorical covariate with $k>2$ categories, since then there is an additional constraint directly on the covariates which is not represented in the quadprog setup. The easiest way to handle this is to set up $\mathbf{X}$ as overparameterized, with $k$ columns for the $k$ categories and take the minimum and maximum per column ignoring the fact that proportions must add up to one. This setup handles the ``summing up"-issue automatically: If $\mathbf{X}_{cat}$ is the $n_j\times k$-submatrix of $\mathbf{X}_j$ corresponding to a dummy-coded categorical covariate with $k$ categories, then $\mathbf{X}_{cat}\cdot \mathbf{1}_k=\mathbf{1}_{n_j}$ and $\mathbf{1}_{n_j}'\mathbf{w}=1$ is one of the constraints of the quadratic optimization setup.

\subsection{Details of the simulations}\label{simproc}

Ten thousand pairs of IPD's are simulated. There are four steps in the process of simulating one pair of IPD:

\begin{enumerate}
\item Generate 15 covariates for IPD A. In particular, we simulate 3 blocks of 5 multi-normal random variables.
\item Simulate another 3 blocks of 5 multi-normal random variables for IPD B. Here, the means for 3 of the 15 covariates are shifted compared to IPD A. Otherwise, the correlation structure remains the same.
\item Simulate a response variable that depends on 6 of the 15 covariates. The dependency is on the underlying continuous scale of these covariates.
\item Dichotomize or categorize 10 of the 15 covariates. For this purpose, thresholds are arbitrarily selected, and are applied the same way to both IPD's.
\end{enumerate}

The R code used are:

\begin{lstlisting}
## step 0 ::: 
## define parameters for multi-normal random variables

n_variables <- 15 ## total number of covariates
block_size <- 5   ## number of covariates in each block

## correlation between covariates within each block
## for block 1, 2, and 3, respectively
rho <- c(0.3, 0.5, 0.7) 

## covariance matrix for the 15 covariates
corr_mat <- diag(1, n_variables, n_variables)
for (i in 1:(n_variables/block_size)) {
  start_index <- (i-1) * block_size + 1
  end_index <- i * block_size
  corr_mat[start_index:end_index, 
           start_index:end_index] <- rho[i]
}
diag(corr_mat) <- 1 ## correct diagonal values

## step 1:::
## simulate covariates for ipd A

n1_obs <- 300 ## number of observations in ipd A

simdt1 <- mvtnorm::rmvnorm(
  n = n1_obs, 
  mean = rep(0, n_variables), 
  sigma = corr_mat
  )

colnames(simdt1) <- paste0('X', 1:15)
simdt1 <- as.data.frame(simdt1) ## convert to dataframe

## step 2 :::
## simulate covariates for ipd B

n2_obs <- 300 ## number of observations in ipd B

simdt2 <- mvtnorm::rmvnorm(
  n = n2_obs, 
  ## shift means for 3 variables
  mean = c(1, 0, 0, 0, 0, ## block 1, X1 shifted
           0, 0, 1, 0, 0, ## block 2, X8 shifted
           0, 0, 0, 0, 1  ## block 3, X15 shifted
  ),
  sigma = corr_mat
  )

colnames(simdt2) <- paste0('X', 1:15)
simdt2 <- as.data.frame(simdt2) ## convert to dataframe

## combine the two datasets together
simdt <- data.frame(rbind(simdt1, simdt2)) %>%
  mutate(study = c(rep('IPD A', n1_obs), 
                   rep('IPD B', n2_obs)))

## step 3: simulate response variable Y :::

simdt <- simdt %>%
    mutate(Y = 0.3 * X1 + 0.2 * X3 + 0.3 * X8 +
               0.1 * X9 + 0.2 * X11 + 0.1 * X15,
           Y = Y + rnorm(n = n1_obs + n2_obs, 
                         mean = 0, sd = 1)
    )

## step 4 :::
## dichotomize or categorize 10 covariates

## variables to dichotomize or categorize
vars_to_categorize <- c("X1", "X2", "X3", 
                        "X6", "X7", "X8", 
                        "X11", "X12", "X13", "X14"
                        )
                        
## threholds
vc.co <- list(
  q.x1 = qnorm(0.238),                    ## x1, binary
  q.x2 = qnorm(0.312),                    ## x2, binary
  q.x3 = qnorm(c(0.12, 0.335, 0.68)),     ## x3, 4 levels
  q.x6 = qnorm(0.439),                    ## x6, binary
  q.x7 = qnorm(0.581),                    ## x7, binary
  q.x8 = qnorm(c(0.23, 0.56)),            ## x8, 3 levels
  q.x11 = qnorm(0.607),                   ## x11, binary
  q.x12 = qnorm(0.712),                   ## x12, binary
  q.x13 = qnorm(0.842),                   ## x13, binary
  q.x14 = qnorm(c(0.18, 0.3, 0.56, 0.72)) ## x14, 5 levels
)

## define function to categorize continuous variables
num2cat <- function(df, vars_to_cat, thresholds){
  for (i in (1 : length(vars_to_cat)) ) {
    df[[vars_to_cat[i]]] <- 
      cut(df[[vars_to_cat[i]]], 
          breaks = c(-Inf, thresholds[[i]], Inf),
          labels = LETTERS[1:(length(thresholds[[i]])+1)])
  }
  return(df)
}

## apply the function 
simdt <- num2cat(df = simdt, 
                 vars_to_cat = vars_to_categorize, 
                 thresholds = vc.co)
\end{lstlisting}

This generates one pair of simulated data. These steps are repeated 10,000 times to simulate the 10,000 pairs of IPD's.

\section{Supplemental material}\label{supps}

\subsection{Summary statistics of one pair of the simulated dataset}\label{summ110}

In Table \ref{tab:summ110}, summary statistics of the 15 covariates of one pair of the simulated IPD's are given. Also presented are weighted means of these covariates after matching using the proposed exact method or the propensity score method.

\begin{table}
\footnotesize
    \centering
    \begin{tabular}{lccccccccccc} \hline
& \multicolumn{2}{c}{} & & \multicolumn{4}{c}{Exact matching weighted means} & & \multicolumn{3}{c}{} \\ 
    & \multicolumn{2}{c}{Observed means} & & \multicolumn{2}{c}{Unconstrained} & \multicolumn{2}{c}{Constrained} & &\multicolumn{3}{c}{Propensity score weighted means} \\ \cline{2-3} \cline{5-8} \cline{10-12} 
    & IPD A & IPD B & & IPD A & IPD B & IPD A & IPD B & & IPD A & IPD B & $|$SMD$|$ \\ \hline
Sample size & $300$ & $300$ & & & & & & & & & \\
ESS & & & & 147.0 & 155.7 & 144.2 & 152.7 & & 98.4 & 62.3 & \\
\multicolumn{12}{l}{X1 (\%)} \\
$\quad$ A & 23.3 & 6.7   & & \multicolumn{2}{c}{11.3} & \multicolumn{2}{c}{11.1} & & 14.5 & 16.3 & 0.050\\ 
$\quad$ B & 76.7 & 93.3 & & \multicolumn{2}{c}{88.7} & \multicolumn{2}{c}{88.9} & & 85.5 & 83.7 & 0.050 \\ 
\multicolumn{12}{l}{X2 (\%)} \\
$\quad$ A & 30.7 & 36.0 & & \multicolumn{2}{c}{33.9} & \multicolumn{2}{c}{33.7} & & 41.0 & 41.6 & 0.014\\ 
$\quad$ B & 69.3 & 64.0 & & \multicolumn{2}{c}{66.1} & \multicolumn{2}{c}{66.3} & & 59.0 & 58.4 & 0.014 \\ 
\multicolumn{12}{l}{X3 (\%)} \\
$\quad$ A & 11.7 & 14.0 & & \multicolumn{2}{c}{12.3} & \multicolumn{2}{c}{13.0} & & 17.7 & 9.4 & 0.244 \\ 
$\quad$ B & 21.3 & 19.0 & & \multicolumn{2}{c}{19.2} & \multicolumn{2}{c}{20.0} & & 19.6 & 20.7 & 0.026 \\ 
$\quad$ C & 38.0 & 38.0 & & \multicolumn{2}{c}{39.7} & \multicolumn{2}{c}{38.0} & & 35.1 & 44.1 & 0.186 \\ 
$\quad$ D & 29.0 & 29.0 & & \multicolumn{2}{c}{28.8} & \multicolumn{2}{c}{29.0} & & 27.5 & 25.8 & 0.040 \\ 
\multicolumn{12}{l}{X4 (mean)} \\
& -0.059 & 0.006 & & \multicolumn{2}{c}{-0.048} & \multicolumn{2}{c}{-0.059} & & -0.131 & -0.161 & 0.030 \\ 
\multicolumn{12}{l}{X5 (mean)} \\
& 0.040  & -0.028 & & \multicolumn{2}{c}{0.008} & \multicolumn{2}{c}{0.008} & & -0.203 & -0.062 & 0.133 \\ 
\multicolumn{12}{l}{X6 (\%)} \\
$\quad$ A & 38.0 & 41.7 & & \multicolumn{2}{c}{35.3} & \multicolumn{2}{c}{38.0} & & 33.8 & 39.4 & 0.117 \\ 
$\quad$ B & 62.0 & 58.3 & & \multicolumn{2}{c}{64.7} & \multicolumn{2}{c}{62.0} & & 66.2 & 60.6 & 0.117 \\ 
\multicolumn{12}{l}{X7 (\%)} \\
$\quad$ A & 54.3 & 58.3 & & \multicolumn{2}{c}{52.8} & \multicolumn{2}{c}{55.8} & & 57.9 & 54.0 &  0.078 \\ 
$\quad$ B & 45.7 & 41.7 & & \multicolumn{2}{c}{47.2} & \multicolumn{2}{c}{44.2} & & 42.1 & 46.0 & 0.078 \\ 
\multicolumn{12}{l}{X8 (\%)} \\
$\quad$ A & 23.0 & 3.7 & & \multicolumn{2}{c}{7.9} & \multicolumn{2}{c}{9.3} & & 13.2 & 12.6 & 0.018 \\ 
$\quad$ B & 30.0 & 15.7 & & \multicolumn{2}{c}{21.3} & \multicolumn{2}{c}{22.8} & & 21.0 & 20.1 & 0.023 \\ 
$\quad$ C & 47.0 & 80.7 & & \multicolumn{2}{c}{70.8} & \multicolumn{2}{c}{67.9} & & 65.8 & 67.3 & 0.032 \\ 
\multicolumn{12}{l}{X9 (mean)} \\
& 0.070 & 0.066 & & \multicolumn{2}{c}{0.171} & \multicolumn{2}{c}{0.070} & & 0.135 & 0.075 & 0.060 \\ 
\multicolumn{12}{l}{X10 (mean)} \\
& 0.122 & -0.004 & & \multicolumn{2}{c}{0.208} & \multicolumn{2}{c}{0.122} & & 0.122 & 0.209 & 0.095 \\ 
\multicolumn{12}{l}{X11 (\%)} \\
$\quad$ A & 64.7 & 62.3 & & \multicolumn{2}{c}{66.9} & \multicolumn{2}{c}{64.3} & & 65.8 & 67.7 & 0.040\\ 
$\quad$ B & 35.3 & 37.7 & & \multicolumn{2}{c}{33.1} & \multicolumn{2}{c}{35.7} & & 34.2 & 32.3 & 0.040 \\ 
\multicolumn{12}{l}{X12 (\%)} \\
$\quad$ A & 72.3 & 73.7 & & \multicolumn{2}{c}{75.5} & \multicolumn{2}{c}{73.7} & & 77.5 & 76.7 & 0.020\\ 
$\quad$ B & 27.7 & 26.3 & & \multicolumn{2}{c}{24.5} & \multicolumn{2}{c}{26.3} & & 22.5 & 23.3 & 0.020 \\ 
\multicolumn{12}{l}{X13 (\%)} \\
$\quad$ A & 84.7 & 88.3 & & \multicolumn{2}{c}{87.1} & \multicolumn{2}{c}{86.3} & & 87.9 & 89.2 & 0.043 \\ 
$\quad$ B & 15.3 & 11.7 & & \multicolumn{2}{c}{12.9} & \multicolumn{2}{c}{13.7} & & 12.1 & 10.8 & 0.043 \\ 
\multicolumn{12}{l}{X14 (\%)} \\
$\quad$ A & 24.0 & 19.3 & & \multicolumn{2}{c}{24.1} & \multicolumn{2}{c}{22.6} & & 20.9 & 30.2 & 0.215 \\ 
$\quad$ B & 11.7 & 11.0 & & \multicolumn{2}{c}{13.8} & \multicolumn{2}{c}{11.7} & & 17.6 & 12.7 & 0.137 \\ 
$\quad$ C & 26.0 & 30.7 & & \multicolumn{2}{c}{28.5} & \multicolumn{2}{c}{27.4} & & 28.0 & 25.3 & 0.061 \\ 
$\quad$ D & 13.7 & 14.0 & & \multicolumn{2}{c}{10.1} & \multicolumn{2}{c}{13.7} & & 9.4 & 10.9 & 0.049 \\ 
$\quad$ E & 24.7 & 25.0 & & \multicolumn{2}{c}{23.5} & \multicolumn{2}{c}{24.7} & & 24.1 & 20.9 & 0.077 \\ 
\multicolumn{12}{l}{X15 (mean)} \\
& -0.011 & 0.946 & & \multicolumn{2}{c}{0.386} & \multicolumn{2}{c}{0.429} & & 0.297 & 0.320 & 0.023 \\ \hline
\end{tabular}
\caption{Summary statistics of one pair of the simulated IPD's} 
\label{tab:summ110}
\end{table}

Table \ref{tab:ybar110} presents the observed and weighted mean responses of the same simulated pair of IPD's. The top two rows are when the response Y depends on the continuous scale of the 6 influencing covariates; the bottom two rows are when the response Y depends on the catigorical scale of four covariates, and the continuous scale of the other two.
\begin{table}[ht]
\centering
\begin{tabular}{lcccccccccc}
  \hline
& \multicolumn{2}{c}{} & & \multicolumn{4}{c}{Exact matching weighted} & & \multicolumn{2}{c}{Propensity score} \\ 
& \multicolumn{2}{c}{Observed means} & & \multicolumn{2}{c}{Unconstrained} & \multicolumn{2}{c}{Constrained} & &\multicolumn{2}{c}{weighted} \\ \cline{2-3} \cline{5-8} \cline{10-11} 
& IPD A & IPD B & & IPD A & IPD B & IPD A & IPD B & & IPD A & IPD B \\ \hline
$\bar{Y}_1,\ \bar{Y}_2$ & -0.028 & 0.684 & & 0.100 & 0.449 & 0.080 & 0.421 & & -0.005 & 0.250 \\
$\bar{Y}_1 - \bar{Y}_2$ & \multicolumn{2}{c}{-0.713} & & \multicolumn{2}{c}{-0.349} & \multicolumn{2}{c}{-0.341} & & \multicolumn{2}{c}{-0.255} \\ \hline
$\bar{Y}_1,\ \bar{Y}_2$ & 0.843 & 1.098 & & 0.898 & 0.921 & 0.887 & 0.903 & & 0.868 & 0.822 \\
$\bar{Y}_1 - \bar{Y}_2$ & \multicolumn{2}{c}{-0.255} & & \multicolumn{2}{c}{-0.022} & \multicolumn{2}{c}{-0.016} & & \multicolumn{2}{c}{0.046} \\ \hline
\end{tabular}
\caption{Mean response Y (observed and weighted) of the pair of IPD's from Table \ref{tab:summ110}} 
\label{tab:ybar110}
\end{table}

\subsection{Additional results}\label{suppsim}

Table \ref{tab:summess} presents summary statistics of the ESS by different matching methods for the 10,000 simulated pairs of IPD's.

\begin{table}[ht]
\centering
\begin{tabular}{lrrrrrr}
  \hline
 & \multicolumn{2}{c}{Unconstrained} & \multicolumn{2}{c}{Constrained} & \multicolumn{2}{c}{PS weight} \\ 
 Statistics & IPD A & IPD B & IPD A & IPD B & IPD A & IPD B \\ 
  \hline
\# of simulations & \multicolumn{2}{c}{10000} & \multicolumn{2}{c}{10000} & \multicolumn{2}{c}{10000} \\ 
Minimum    &  97.2   & 104.7   &  95.6   & 102.5   &   1.10   &   1.10   \\ 
1st quartile & 134.6   & 142.2   & 132.8   & 140.3   &  51.70   &  43.48   \\ 
Median  & 141.3   & 149.3   & 139.7   & 147.6   &  82.40   &  78.00   \\ 
Mean    & 141.3   & 149.3   & 139.7   & 147.5   &  81.95   &  79.33   \\ 
3rd quartile & 148.3   & 156.7   & 146.8   & 155.0   & 110.90   & 112.72   \\ 
Maximum    & 184.8   & 191.3   & 184.1   & 190.8   & 208.10   & 207.50   \\ 
Standard deviation & 10.21 & 10.74 & 10.37 & 10.94 & 39.46 & 44.16 \\ 
NA's & \multicolumn{2}{c}{0} & \multicolumn{2}{c}{46} & \multicolumn{2}{c}{0} \\ \hline

\end{tabular}
\caption{Summary statistics of ESS for the 10,000 pairs of simulated IPD's} 
\label{tab:summess}
\end{table}

Table \ref{tab:summYbar_num} presents the summary statistics of the mean response Y (observed and weighted). In this table, Y is simulated to depend on the underlying continuous scale of the 6 influencing covariates.

\begin{table}[ht]
\centering
\begin{tabular}{lrrrr}
  \hline
  & & \multicolumn{3}{c}{Weighted difference} \\ \cline{3-5}
  Statistics & Observed & Unconstrained & Constrained & Propensity score \\ 
  \hline
\# of simulations & 10000 & 10000 & 10000 & 10000 \\ 
Minimum    & -1.051   & -0.724   & -0.733   & -1.920   \\ 
1st quartile & -0.766   & -0.372   & -0.367   & -0.445   \\ 
Median  & -0.700   & -0.289   & -0.284   & -0.308   \\ 
Mean    & -0.699   & -0.288   & -0.283   & -0.277   \\ 
3rd quartile & -0.633   & -0.203   & -0.199   & -0.154   \\ 
Maximum    & -0.275   &  0.151   &  0.152   &  2.548   \\ \hline
Standard deviations & 0.097 & 0.124 & 0.125 & 0.279 \\ 
NA's    & 0 & 0 & 46   & 0 \\  \hline
\end{tabular}
\caption{Summary statistics of the mean differences in the response variable $Y$ among the 10,000 pairs of IPD's. $Y$ depends on the underlying continuous scale of 6 covariates.} 
\label{tab:summYbar_num}
\end{table}

Table \ref{tab:summYbar_cat} presents the summary statistics of the mean response Y (observed and weighted). In this case, Y is simulated to depend on the categorical scale of four covariates, and the continuous scale of the other two.

\begin{table}[ht]
\centering
\begin{tabular}{lrrrr}
  \hline
  & & \multicolumn{3}{c}{Weighted difference} \\ \cline{3-5}
  Statistics & Observed & Unconstrained & Constrained & Propensity score \\ 
  \hline
\# of simulations & 10000 & 10000 & 10000 & 10000 \\ 
Minimum    & -0.559   & -0.408   & -0.412   & -1.801   \\ 
1st Quartile & -0.301   & -0.080   & -0.081   & -0.146   \\ 
Median  & -0.2411   & -0.0015   & -0.0012   & -0.0185   \\ 
Mean    & -0.2405   &  0.0002   &  0.0003   & -0.0041   \\ 
3rd Quartile & -0.182   &  0.0810   &  0.082   &  0.111   \\ 
Maximum    &  0.108   &  0.422   &  0.460   &  1.962   \\ \hline
Standard deviation & 0.087 & 0.118 & 0.119 & 0.236 \\  
NA's    & 0 & 0 & 46   & 0 \\  \hline
\end{tabular}
\caption{Summary statistics of the mean difference in the response variable $Y$ among the 10,000 pairs of IPD's. $Y$ depends on the categorical scale of four covariates, and the continuous scale of the other two.} 
\label{tab:summYbar_cat}
\end{table}

\begin{figure}
    \centering
    \includegraphics[width=1\textwidth]{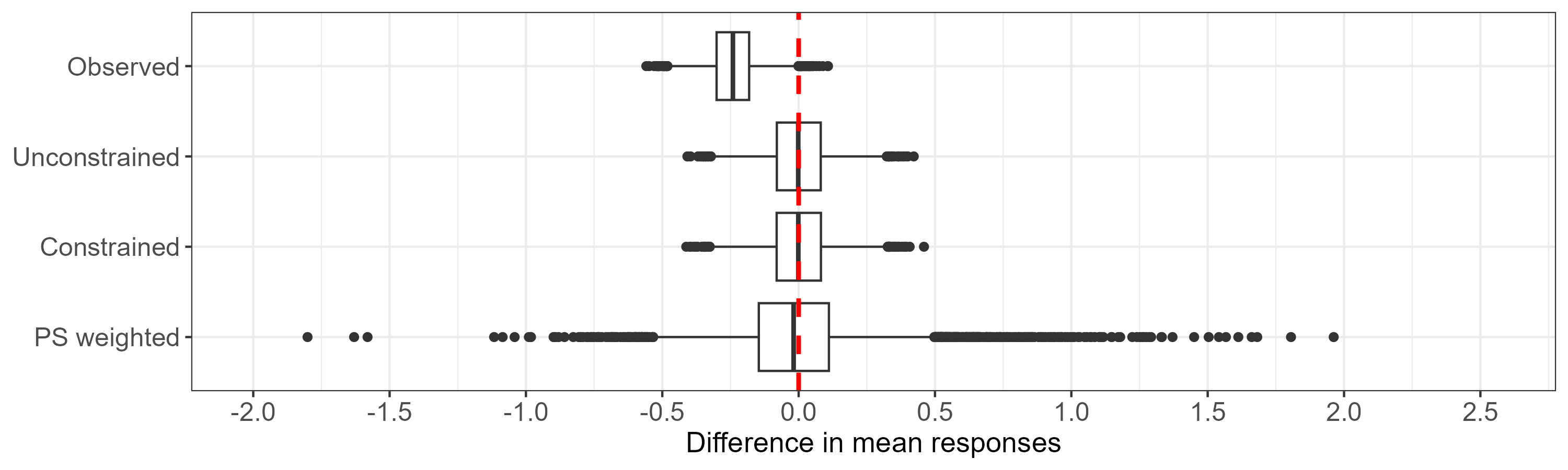}
    \caption{Boxplots of the mean difference in the response variable $Y$ among the 10,000 pairs of IPD's. $Y$ depends on the categorical scale of 4 covariates, and the continuous scale of the other two.}
    \label{fig:boxYbar_cat}
\end{figure}

Figure \ref{fig:boxYbar_cat} is the boxplot of the summary statistics in Table \ref{tab:summYbar_cat}, where the response variable $Y$ depends on the categorical scale of four covariates, and the continuous scale of the other two.

\end{document}